# Improving the critical temperature of MgB$_2$ superconducting metamaterials induced by electroluminescence

Zhiwei Zhang, Shuo Tao, Guowei Chen, Xiaopeng Zhao*

Smart Materials Laboratory, Department of Applied Physics, Northwestern Polytechnical University, Xi'an 710129, PR China

*Corresponding author: Prof. Xiaopeng Zhao

Tel.: +86-29-8843-1662

Fax: +86-29-8849-1000

E-mail: xpzhao@nwpu.edu.cn

**Abstract**

The MgB$_2$ superconductor was doped with electroluminescent Y$_2$O$_3$:Eu, to synthesise a superconducting metamaterial. The temperature dependence of the resistivity of the superconductor indicates that the critical temperature ($T_C$) of samples decrease when increasing the amount of doped Y$_2$O$_3$ nanorods, due to impurity (Y$_2$O$_3$, MgO and YB$_4$). However, the $T_C$ of the samples increase with increasing amount of doped Y$_2$O$_3$:Eu$^{3+}$ nanorods, which are opposite to doped Y$_2$O$_3$ nanorods. Moreover, the transition temperature of the sample doped with 8 wt. % Y$_2$O$_3$:Eu$^{3+}$ nanorods is higher than those of doped and pure MgB$_2$. The $T_C$ of the sample doped with 8 wt. % Y$_2$O$_3$:Eu$^{3+}$ nanorods is 1.15 K higher than that of the sample doped with 8 wt. % Y$_2$O$_3$. The $T_C$ of sample doped with 8 wt. % Y$_2$O$_3$:Eu$^{3+}$ is 0.4 K higher than that of pure MgB$_2$. Results indicate that doping electroluminescent materials into MgB$_2$ increases the transition temperature; this novel strategy may also be applicable to other superconductors.

**Introduction**

The binary metal boride superconductor $MgB_2$ has attracted considerable attention in theoretical studies and applications [1-6]. $MgB_2$ presents great potential in superconductive device applications because of its advantages relatively high critical temperature ($T_C$ nearly 40 K) [1], long superconducting correlation length [7], high critical current density, wide energy gap [8] and facile preparation.

$MgB_2$ is an inexpensive Type II superconductor inexpensive with a simple structure and can be easily synthesised [3]. However, the application of $MgB_2$ is restricted by its lower $T_C$ compared with those of high-temperature superconductors. Several studies attempted to increase $T_C$ by using doping agents. Results indicated that dopants decrease the $T_C$ of the $MgB_2$ superconductor [9] and affect its physical properties, such as carrier concentration, lattice constant and crystallinity [10]. For example, doping aluminium (or carbon) in place of Mg (or B) in $MgB_2$ significantly decreases the $T_C$ of $MgB_2$ [11]. Recently, Ma et al. recently increased the $T_C$ of commercial $MgB_2$ from 33.0 K to 37.8 K through nonsubstitutional hole-doping of the $MgB_2$ structure with small, single-wall carbon nanotube inclusions [12].

Metamaterials have grained increased attention because of their special properties and application potential [13-17]. In 2007, our group proposed that combining inorganic electroluminescent (EL) materials with metamaterials can induce substantial change in superconducting materials, left-handed materials, photonic crystals and so on [18]. In the same year, Jang et al studied the effects of ZnO doping on the superconductivity and crystal structure of the (Bi, Pb)-2233 superconductor [19].

Light-induced superconductivity has recently been considered as a research hotspot. In 2011, Cavalleri et al. used mid-infrared femtosecond pulses to transform a stripe-ordered compound, namely, nonsuperconducting $La_{1.675}Eu_{0.2}Sr_{0.125}CuO_4$, into a transient three-dimensional superconductor [20]. In 2013, Cavalleri et al. experimentally demonstrated the excitation of Josephson plasma in $La_{1.84}Sr_{0.16}CuO_4$, by using intense narrowband radiation from infrared free-electron laser tuned to the 2-THz Josephson plasma resonance [21]. The group believed that laser pulse led to deformed crystal structures and induced superconductivity [22].

$Y_2O_3:Eu^{3+}$ phosphor is a well-known red phosphor used in fluorescent light, field emission displays and cathode ray tubes, because of its excellent luminescence efficiency, narrow spectra, high brightness and environmental and chemical stability [23, 24]. In this paper, we examined the effect of electroluminescence on $T_C$ by doping $Y_2O_3$, $Y_2O_3:Sm^{3+}$ and $Y_2O_3:Eu^{3+}$ nanorods into the $MgB_2$ superconductor to produce a superconducting metamaterial. The appropriate doping content for increasing the $T_C$ of the $MgB_2$ superconducting metamaterial was demonstrated by controlling the concentration of $Y_2O_3:Eu^{3+}$ nanorods.

**Experimental Section**

Bulk $MgB_2$ sample was synthesised with $Y_2O_3:Eu^{3+}$ nanorods by traditional solid-state sintering [25]. Briefly, 0.22 g of magnesium (Mg, 97% purity, 10 μm in size), 0.18 g of boron (B, 99.99% purity, 1 μm in size) and $Y_2O_3:Eu^{3+}$ (nanorods, 1-2 μm in length, 150-200 nm in diameter) were uniformly mixed at different mass ratios and ground for 20 min in an agate mortar. The mixture was transferred into a mould and pressed into cylindrical tablets at 20 MPa for 30 min. The samples were then sintered at 850 ℃ for 2 h under flowing high-purity Ar gas (99.99 % purity) at a heating rate of 5 ℃ min$^{-1}$. These procedures were also used toprepare MgB$_2$ doped with Y$_2$O$_3$ or Y$_2$O$_3$:Sm nanorods. Afterwards, the crystal structure was investigated through X-ray diffraction (XRD) by using a BRUKER D8 Advanced X-ray diffractometer with Cu Kα irradiation. Resistivity was measured using the four-probe technique. Electric field was imported through the electrodes. The measuring current was 100 mA.

**Results and discussion**

The image in Fig. 1 shows the X-ray diffraction patterns of the series of MgB$_2$ doped with different amounts of Y$_2$O$_3$ and Y$_2$O$_3$: Eu$^{3+}$ nanorods. Similar with other reports, the XRD results of pure MgB$_2$ indicate that the MgO phase exists in MgB$_2$. In comparison, the XRD results of the doping samples show that YB$_4$ phase, which is generated from the reaction of B and Y$_2$O$_3$, appears in the other samples. Surplus Y$_2$O$_3$ presents in XRD patterns with increasing amount of doping material. The XRD patterns also indicate that samples doped with Y$_2$O$_3$ nanorods are similar to those doped with Y$_2$O$_3$: Eu nanorods, because of the effective incorporation of Eu into the lattice of Y$_2$O$_3$.

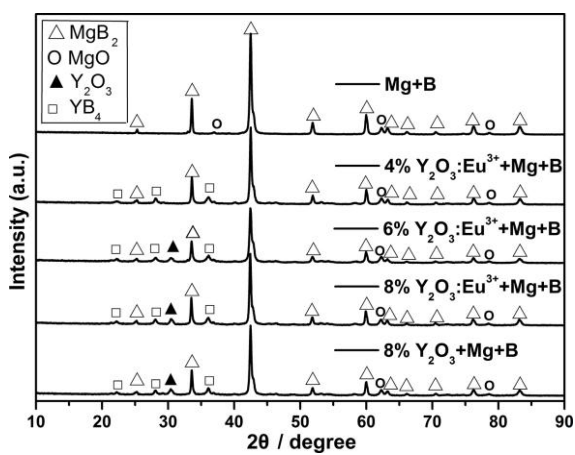

**Fig. 1** X-ray diffraction patterns of $Y_2O_3$ and $Y_2O_3$:$Eu^{3+}$-doped $MgB_2$ samples after sintering.

This study aims to prepare a composite superconducting metamaterial consisting of $MgB_2$ superconductor and $Y_2O_3$:Eu electroluminescent material. Firstly, we synthesised $Y_2O_3$:Eu nanorods on the basis of relevant research about $Y_2O_3$:Eu powders [26]. The Fig. 2 shows the EL spectrum of $Y_2O_3$:Eu nanorods. The spectrum indicates that the strongest peak centred at 613 nm corresponds to $Eu^{3+}$ ions typical transition of from $^5D_0$ to $^7F_2$. Moreover, the full width at half maximum of the main emission peak is only 7 nm, suggesting superior monochromaticity, which is consistent with other reports. Secondly, the superconducting metamaterials was prepared by doping $Y_2O_3$:Eu nanorods into $MgB_2$. Electric field in bulk $MgB_2$ stimulates the electroluminescence of $Y_2O_3$:Eu. To further demonstrate the effect of electroluminescence in $MgB_2$, we synthesised $Y_2O_3$:Sm nanorods whose electroluminescence intensity is much lower than that of $Y_2O_3$:Eu (Fig. 2).

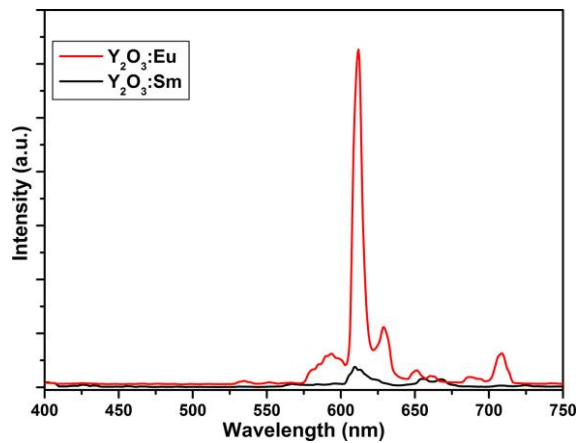

**Fig. 2** Electroluminesence spectrum of $Y_2O_3$:Eu nanorods and $Y_2O_3$:Sm nanorods.

Fig. 3 presents the temperature dependence of the resistivity of the series of $MgB_2$ samples doped with different amounts of $Y_2O_3$ or $Y_2O_3$:$Eu^{3+}$ nanorods. As is observed in Fig. 3(a), the superconducting $T_C$ of pure $MgB_2$ is 37 K, which is higher than those of other doped samples. The transition width of pure $MgB_2$ is only 0.4 K, which is also the minimum value in all samples. The superconducting $T_C$ decreases, whereas the width of transition increases with increasing doping amount. For instance, the $T_C$ of the sample doped with 8 wt. % $Y_2O_3$ is 36.25 K, which is lower than that of pure samples. The presence of impurities, such as $Y_2O_3$, MgO and $YB_4$, induced the deterioration of intercrystalline connectivity [27, 28].

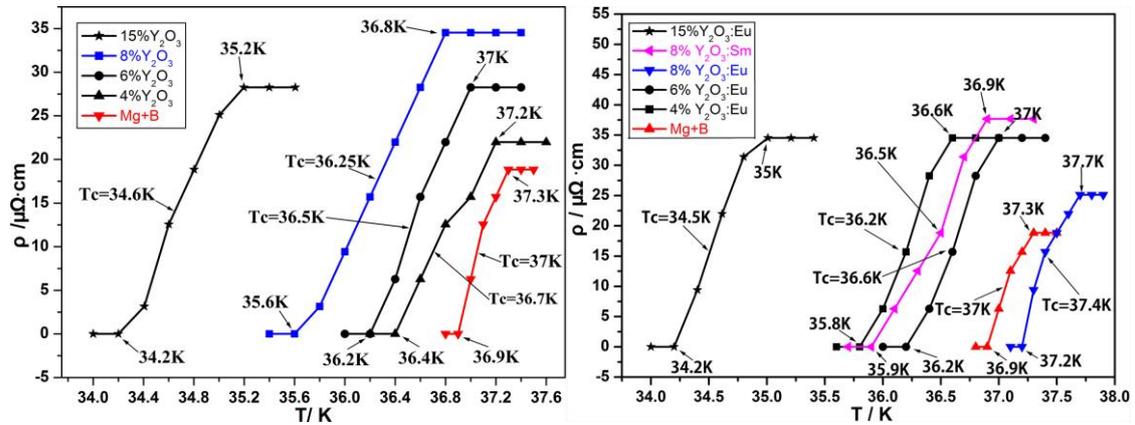

**Fig. 3** Temperature dependence of resistivity of the undoped, 4, 6 and 8 wt.% doped $MgB_2$ samples. (a) Doped $Y_2O_3$ nanorods; (b) doped $Y_2O_3$:$Eu^{3+}$ nanorods.

The image in Fig. 3(b) shows that the $T_C$ of samples doped with 4 wt. % or 6 wt. % $Y_2O_3$:$Eu^{3+}$ nanorods is lower than that of pure $MgB_2$, which is consistent with those of samples doped with $Y_2O_3$. However, the $T_C$ of sample doped with 8 wt. % $Y_2O_3$:$Eu^{3+}$ nanorods is the highest among all samples, including pure $MgB_2$. The $T_C$ of sample doped with 8 wt. % $Y_2O_3$:$Eu^{3+}$ nanorods is 1.15 K higher than that of the sample doped with 8 wt. % $Y_2O_3$ nanorods. It's peculiar that the $T_C$ of sample doped with 8 wt. % $Y_2O_3$:$Eu^{3+}$ nanorods is 0.4 K higher than that of pure $MgB_2$. The curve of $MgB_2$ doped with 8 wt. % $Y_2O_3$:Sm is also shown in Fig. 3b to confirm whether electroluminescence or rare earth element affect the transition temperature. The transition temperature of MgB$_2$ doped with 8 wt. % Y$_2$O$_3$:Sm is 0.9 K lower than that of MgB$_2$ doped with 8 wt. % Y$_2$O$_3$:Eu (Fig. 3b). Europium, which results in bright red light, is critical for $T_C$.

The images in Fig. 4 presents the $T_C$ of the series of MgB$_2$ samples doped with different amounts of Y$_2$O$_3$ or Y$_2$O$_3$:Eu$^{3+}$ nanorods. The $T_C$ values of samples decrease with increasing amount of doped Y$_2$O$_3$ nanorods, which is consistent with previous reports. However, it's fantastic that the $T_C$ values of samples increase with the amount of doped Y$_2$O$_3$:Eu$^{3+}$ nanorods. Additionally, the $T_C$ of sample doped 8 wt. % Y$_2$O$_3$:Eu$^{3+}$ nanorods is 0.4 K higher than that of pure MgB$_2$. However, the transition temperature decreases rapidly when the doping percentage of Y$_2$O$_3$:Eu is larger than 8. The phenomenon cannot be explained by existing theories. The mechanism of increasing superconducting $T_C$ still needs further studies. The transition temperature is one of the most crucial factors that limit the application of superconductors. As such, increasing the transition temperature of superconductors is of great interest. Hence, doping EL materials in MgB$_2$ is a novel strategy for increasing the transition temperature and may also be applicable to other superconductors.

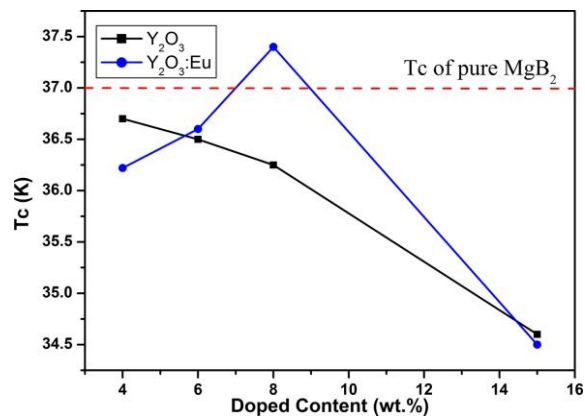

**Fig. 4** The $T_C$ of the series MgB$_2$ samples doped different content of Y$_2$O$_3$ or Y$_2$O$_3$:Eu$^{3+}$ nanorods.

**Conclusion**

We synthesised a composite material, MgB$_2$ superconductor doped with the EL materials Y$_2$O$_3$:Eu, which is called superconducting metamaterial. The XRD results indicate that the crystallinity and the purity of MgB$_2$ decrease with increasing the doping content. However, more Y$_2$O$_3$:Eu$^{3+}$ would remain in the samples at a high doping content. The temperature dependence of resistivity indicate that the $T_C$ values of samples decrease with increasing amount of doped Y$_2$O$_3$ nanorods; this finding is consistent with previous reports. However, it's remarkable that the $T_C$ values of samples increase with increasing amount of doped Y$_2$O$_3$:Eu$^{3+}$ nanorods. The $T_C$ of the sample doped with 8 wt. % Y$_2$O$_3$:Eu$^{3+}$ is 1.15 K higher than that of sample doped with 8 wt. % Y$_2$O$_3$. It's remarkable that the $T_C$ of the sample doped with 8 wt. % Y$_2$O$_3$:Eu$^{3+}$ is 0.4 K higher than that of pure MgB$_2$. However, the transition temperature of the sample doped with 8 wt. %Y$_2$O$_3$:Sm nanorods is 0.9 K lower than that of the sample doped with 8 wt. %Y$_2$O$_3$:Eu nanorods. Transition temperature is one of the most crucial factors that limit the application of superconductors. Doping EL materials in MgB$_2$ is a novel strategy for increasing the transition temperature and may also be applicable to other superconductors.

**Acknowledgements**

This work was supported by the National Natural Science Foundation of China for Distinguished Young Scholar under Grant No.50025207.